\documentclass[aps,pra,11pt,tightenlines,amsmath,amssymb,amsfonts,nofootinbib,showpacs,notitlepage]{revtex4-1}
\usepackage{epsfig}
\usepackage{amsmath}
\usepackage{amssymb}
\usepackage{amsthm} 
\usepackage{bm}
\usepackage{color}

\newcommand{\bea}{\begin{eqnarray}}
\newcommand{\ea}{\end{eqnarray}}
\newcommand{\eea}{\end{eqnarray}}

\newcommand{\sumint}[1]

\begin{document}

\title{Verifying the observer dependence of quasiparticle counts 
in the analogue gravity of dilute ultracold quantum gases}

\author{Uwe R. Fischer}

\affiliation{Seoul National University, Department of Physics and Astronomy \\  Center for Theoretical Physics, 
151-747 Seoul, Korea}

\begin{abstract}
\end{abstract}


\maketitle


\section{The relativity of the quasiparticle content of a quantum field}

The particle content of a quantum field in flat or curved space-time depends on the motional state of the observer, on the way the observer's particle detector couples to the quantum field, and on the frequency standard in which the detector carried by the observer measures the quanta to be detected.
A particular manifestation of this observer dependence is the Unruh effect, which consists in the fact that a constantly accelerated detector in the Minkowski vacuum responds as if it were placed in a thermal bath with temperature proportional to its acceleration 
\cite{unruh76,BirrellDavies,fulling,Luis}. 
The deceptively simple fact that the particle content of a quantum field, measured by a suitable detector 
attached to the observer in a supposedly empty Minkowski spacetime, depends on the motional state of the observer  has eluded direct observation so far.
The value of the Unruh temperature 
$ %
T_{\rm Unruh} = 
[\hbar /(2\pi k_ {\rm B} c_{\rm L})]  a = 4\, {\rm K} \times a[10^{20} g_{\oplus}] \,,
$ 
where $a$ is the acceleration of the detector in Minkowski space 
($g_{\oplus}$ is the gravity acceleration on the surface of the Earth), and 
$c_{\rm L}$ the speed of light, makes it obvious that an observation of the 
effect is a less than trivial undertaking. Proposals for a measurement with ultraintense short pulses of electromagnetic radiation have been put forward in, e.g., Refs. \cite{Yablonovitch,chenUnruh,Habs}, cf. for an extensive discussion  and more references the review \cite{Luis}.
Apart from the smallness of the effect due to a proportionality constant containing the ratio of Planck constant divided by Boltzmann constant times speed of light, the tiny thermal radiation signal needs to be detected on top of a (in comparison) huge background signal and in the potential presence of 
dissipative effects.

In analogue gravity \cite{unruh,BLVReview,Grisha,GrishaTed,Matt,SchuetzUnruh,SlowLight,PGRiemann}, the excitations  above a given vacuum, propagating in a generally curved spacetime, are quasiparticles like, e.g.,  
phonons, antiferromagnetic magnons, and  photons in nonlinear dielectric media. 
An analogue gravity 
setup of observer-related quantum-field-theoretical phenomena has one distinct advantage over real, i.e.\,\,Einsteinian, gravity: 
there exists the preferred reference frame ``laboratory'', which allows
to state in an absolute manner whether actual dissipation has taken place, that is whether the quasiparticles created are due to the choice of the observer only, or whether some friction effect creates them, slowing down the 
motion of the system under consideration as observed by the experimentalist in the lab.
Two archetypical examples for the latter dissipative process are cosmological particle creation due to nonadiabatic
effects (squeezing of the vacuum) and Hawking radiation from black hole horizons. 
This sets the purely observer-related Unruh effect apart from dissipative effects in an operational, that 
is measurable manner. 
At the same time, the existence of the absolute reference frame laboratory represents a certain limitation of 
analogue gravity, because it implies the (re-)introduction of an ``ether''.
While an ether is to some extent implicit in any quantum field theory, which is ultimately a 
theory of quasiparticles above some vacuum \cite{Wilczek}, the ether in question here introduces a preferred
``absolute'' time coordinate (measured by a clock ticking in the lab frame), 
a concept alien to 
relativity.

It rapidly became apparent that the simulation of 
effective space-times can be very efficiently and with high precision 
carried out in Bose-Einstein condensates (BECs), where the relevant quasiparticles are phonons \cite{CSM}. BECs  offer the primary advantages of dissipation-free superflow, high controllability of the physical parameters involved, and the accessibility of ultralow temperatures \cite{Picokelvin}. Analogues of the dissipative phenomena of cosmological particle creation and Hawking radiation have been discussed extensively in the literature by now cf., e.g. \cite{BLVPRA,0303063,Micha,Prain} for cosmological particle creation and 
 \cite{Macher,Balbinot,Recati} for Hawking radiation by sonic black holes \cite{MattHawking,Garay}. 
As regards the experimental realization of such ``dumb'' 
holes in BECs, some progress has been made as well \cite{Lahav}. 
The nondissipative Gibbons-Hawking effect \cite{Gibbons}, which is the analogue of the Unruh effect for a freely falling observer in a de Sitter spacetime \cite{de Sitter}, will
be discussed in what follows in its analogue gravity 
implementation for  BECs \cite{PRL,PRD}. For the Unruh effect itself, some aspects of its possible realization in BECs, using the same type of detector as first employed in the latter two references, have been put forward in \cite{Retzker}. We note that the experimental sensitivity for detecting phonons in ultracold dilute quantum gases can reach down to the few or even single-phonon level, which is a basic requirement for detecting the subtle effects of quantum radiation phenomena \cite{RalfPRL06}.  
We also note that an alternative setup to detect the analogue Gibbons-Hawking effect 
has been discussed in ion traps \cite{Menicucci} 
(as well as cosmological particle creation and Hawking radiation, see \cite{Petersen,Horstmann}).
While the analogy to an expanding universe does not hold there, as there 
is no real (spatial) expansion of the ion array taking place, 
the advantage of such an ion trap setup
is that the measurement, because of the spatial stationarity of the central ion locations, 
be made over a comparatively long time.
The modulation of the laser coupling electronic states of the ions and their
vibrational motion (constituting the phononic quasiparticles to be detected), there encodes 
the necessary exponential relation between detector proper time and de Sitter time.

\section{Expanding Bose-Einstein condensates}\label{ExpandBEC}

In a seminal paper for the very idea of analogue gravity to be established, it was 
shown by Unruh that the action of the phase fluctuations $\Phi$  in a moving 
inhomogeneous superfluid may be written in the form \cite{unruh}, also see \cite{Matt}
(we set $\hbar = m =1$, where $m$ is the mass of a superfluid 
constituent particle): 
\begin{eqnarray}
S  & = & \int d^{D+1}x 
\frac{1}{2g} \left[ -\left(\frac\partial{\partial t} \Phi -{\bm v} 
\cdot \nabla \Phi\right)^2 + c^2  (\nabla \Phi)^2 \right] \nonumber\\
& \equiv & \frac12 \int d^{D+1}x 
\sqrt{-{\sf g}} {\sf g}^{\mu\nu} 
\partial_\mu \Phi \partial_\nu \Phi \,. \label{action}
\end{eqnarray}
Here, $\bm v({\bm x},t)$ is the superfluid background velocity, 
$c({\bm x},t)=\sqrt{g\rho_0({\bm x},t)}$ is the velocity of sound, where
$g$ is a constant describing the interaction between the constituent
particles in 
the superfluid ($1/g$ is the compressibility of the fluid), 
and $\rho_0({\bm x},t)$ is the background density.
In the second line of (\ref{action}), the conventional 
hydrodynamic action is identified with the action 
of a minimally coupled scalar field in an effective space-time which is generally curved \cite{PGRiemann}. 
Furthermore, the velocity potential of the sound perturbations in the BEC 
satisfies the canonical
commutation relations of a relativistic scalar field \cite{SlowLight}.
The theory of phononic 
quasiparticles in the inhomogeneous BEC is thus kinematically identical to that 
of a massless scalar field propagating on the background 
of curved space-time in $D+1$ dimensions. We therefore have the exact mapping, on this level of kinematics, of 
the equation of motion for phononic quasiparticles in a nonrelativistic superfluid,  
to quantized massless scalar fields propagating on a curved space-time  
background with local Lorentz invariance. 
While some dynamical aspects of gravity like backreaction and the cosmological constant issue have been studied
as well in BECs, see e.g. \cite{BR,LiberatiPRL}, the actual Einsteinian gravity part of the action, proportional to the Ricci curvature scalar, is suppressed by the small gas parameter $D_p = ({\rho_0 a_s^3})^{1/2}$ 
measuring the diluteness of the system \cite{MattMPL,LHY}, where $a_s$ is the scattering length characterizing atomic  collisions in the dilute and ultracold BEC, related to the coupling constant by $g=4\pi a_s$.
Hence Einstein gravity will be subdominant in determining the background metric as long as the gas is
dilute with respect to the interaction range.

\begin{center}
\begin{figure}[bt]
\vspace*{1em}
\includegraphics[width=0.8\textwidth]{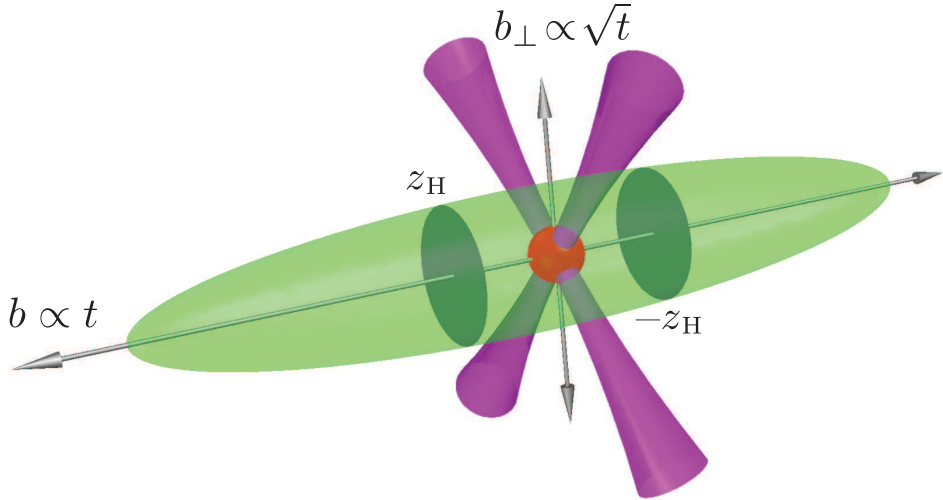}
\caption{\label{Fig1} Expansion of a cigar-shaped Bose-Einstein 
condensate as the sonic analogue of a 1+1D de Sitter universe.
The violet surfaces represent laser beams creating a tight optical potential 
well in the center, hosting the two-level system 
Atomic Quantum Dot (red), 
acting as the detector. The dark surfaces at symmetric axial locations $z_{\rm H}$ and $-z_{\rm H}$ represent the sonic horizon. 
The scaling of the expansion rates along axial and perpendicular directions with time, $b(t)$ and $b_\perp(t)$,   
are derived in the text.}
\end{figure}
\end{center}

To model effective curved space-times for quasiparticles, 
we will consider the evolution of the BEC if we change the harmonic 
trapping, keeping the BEC in a well-defined region of space in the lab, with time.   
For a description of the expansion (or contraction) of a BEC, the so-called scaling 
solution approach is conventionally used  \cite{Scaling}. 
One starts from a cigar-shaped BEC containing a large number of constituent particles, i.e., which is in the 
so-called Thomas-Fermi (TF) limit, which in BEC terminology eliminates any gradients of the 
condensate wave function  \cite{BaymPethick}.  
According to \cite{Scaling}, the evolution
of a Bose-condensed atom cloud under temporal
variation of the trapping frequencies 
$\omega_\parallel(t)$ and $\omega_{\perp}(t)$ 
(in the axial and radial directions, 
respectively) can then be described by the following 
solution for the condensate wave function
\begin{equation}
\Psi = \frac{\Psi_{\rm TF}}{b_\perp {\sqrt b}} 
\exp\left[-i \int g \rho_0 ({\bm x}=0,t) dt+ i\frac{\dot b z^2}{2b}  
+ i\frac{\dot b_\perp r^2}{2b_\perp} 
\right] . \label{Wavefunction}
\end{equation}
Here, $b_{\perp}$ and $b$ are the scaling parameters describing the condensate
evolution in the radial ($\hat r$) and axial ($\hat z$) directions, 
cf.\,\,Fig.\,\ref{Fig1}.
The  initial ($b=b_\perp=1$) mean-field condensate density is given by
the usual TF expression  
\begin{equation}
|\Psi_{\rm TF}|^2=\rho_{\rm TF}(r,z)
=\rho_m \left(1-\frac{r^2}{R_{\perp}^2}-\frac{z^2}{R_\parallel^2}\right).
\end{equation}
Here, $\rho_m$ is the maximum density (in the center of the cloud) and 
the squared initial TF radii are $R^2_\parallel= 2\mu/\omega_\parallel^2 $ 
and $R^2_{\perp}=2\mu/\omega_{\perp}^2$.
The initial chemical  potential $\mu=\rho_m g$. In our cylindrical 3D trap, 
we have for the initial central density 
$$
\rho_m = 
\left(\frac{6 N \omega_\perp^2 \omega_\parallel}{\sqrt 8 \pi g^{3/2}}\right)^{2/5}
.
$$
The condition that TF be valid implies that 
$\mu\gg\omega_\parallel,\omega_{\perp}$. 
The solution (\ref{Wavefunction}) of the Gross-Pitaevski\v\i\/ mean-field
equations becomes exact in this TF limit, independent of the ratio 
$\omega_\parallel/\omega_\perp$. 
However, the solution becomes exact also in the limit that 
$\omega_\perp/\omega_\parallel \rightarrow 0$, 
independent of the validity of the TF limit, the system then 
acquiring an effectively two-dimensional character 
\cite{PitaRosch,DalibardBreathing}. 
We will see below that in the limit  of a very long cigar,  
$\omega_\parallel/\omega_\perp \rightarrow 0$, there is an 
``adiabatic basis'' in which no axial excitations are created 
during the expansion. With respect to that basis there are, 
in particular, no unstable solutions possible, implying the 
stability of the expanding gas against perturbations. 

According to (\ref{Wavefunction}), the condensate density evolves as
\begin{equation}
\rho_0(r,z,t)=\frac{\rho_{\rm TF}\left({r^2}/{b_{\perp}^2},{z^2}/{b^2}\right)}
{b_{\perp}^2b}, \label{scaleddensity} 
\end{equation}
and the superfluid velocity
\begin{equation}
{\bm v}=\frac{\dot b_{\perp}}{b_{\perp}} r {\bm e}_r +\frac{\dot b}{b}z {\bm e}_z.\label{scaledv}
\end{equation}
is the gradient of the condensate  phase in 
Eq. (\ref{Wavefunction}). It increases linearly to the axial and 
radial boundaries of the condensate. 

\vspace*{-1em}
\vspace*{-1em}

The excitations in the limit 
$\omega_{\parallel}/\omega_\perp \rightarrow 0$  
were studied in \cite{Zaremba}.
The description of the modes is based 
on an adiabatic separation for the axial and
longitudinal variables of the phase fluctuation field: 
\begin{equation}
\Phi (r,z,t) = \sum_n \phi_n(r) \chi_n(z,t), \label{sepansatz}
\end{equation}
where $\phi_n(r)$ is the radial wavefunction characterized by the 
quantum number $n$ (we consider only zero angular momentum modes). 
The above ansatz incorporates the fact that for strongly elongated traps the dynamics of the condensate
motion separates into a fast radial motion and a slow axial motion, which 
are essentially independent. 
The $\chi_n(z,t)$ are the mode functions for travelling wave solutions in the 
$z$ direction (plane waves for a condensate at rest read 
$\chi_n \propto \exp [- i \epsilon_{n,k} t + kz]$). The radial motion 
is assumed to be ``stiff'' such that the radial part is effectively 
time independent, because the radial time scale for adjustment of the 
density distribution after a perturbation is much less than the axial 
oscillation time scales of interest.  
The ansatz (\ref{sepansatz}) 
works independent from the ratio of healing length and 
radial size of the BEC cigar.  In the limit that the healing length is much less
than the radial size, TF wave functions are used, in the opposite limit, 
a Gaussian ansatz for the radial part of the wave function $\phi_n(r)$
is appropriate.  

For axial  excitations characterized by a wavelength $\lambda=2\pi/k$ 
exceeding the radial size $R_\perp$ of the condensate, we have $kR_\perp \ll 1$, and 
the dispersion relation reads, in the TF limit for the radial wave function
\cite{Zaremba} 
\begin{eqnarray}\label{zarembadisprel}
\epsilon^2_{n,k} 
& = & 2\omega_{\perp}^2 n (n+1)+c_0^2 k^2, 
\end{eqnarray}
where $c_0 =\sqrt{\mu/2} $ and the chemical potential $\mu = \omega_{\perp}^2 R_{\perp}^2/2$. 
Observe that the central speed of sound $c_0$ 
is reduced by a factor $\sqrt2$ from 
the  well-established value $\sqrt{g\rho_m}=\sqrt\mu$ 
for an infinitely extended liquid \cite{Zaremba}.  

The Eqs. (\ref{sepansatz}) and (\ref{zarembadisprel}) can be 
generalized for an expanding condensate. 
Substituting in Eq.\,(\ref{sepansatz}) the rescaled radial wavefunction 
$\phi_n\equiv\phi_n(r/b_\perp)$, and inserting the result 
into the action (\ref{action}), integrating
over the radial coordinates, we find the following 
effective action for the axial modes of a given radial quantum number $n$:
\begin{eqnarray}
S_n  &= & \int dt  dz \frac{b_\perp^2 C_n(z)}{2g}
\left[ -\left(\partial_t\chi_n -{v}_z \partial_z\right)^2 
+ \bar c^2_n(z) (\partial_z \chi_n)^2 
+ M^2_n(z)\chi_n^2 \right],
\label{1daction}
\end{eqnarray}
where the common ``conformal'' factor $C_n(z)$ is given by  
\begin{equation}
b_\perp^2 C_n(z)=\int_{r<r_m} d^2r \phi^2_n . 
\end{equation}
The integration limits are fixed by the $z$ dependent 
radial size of the cigar 
$r^2_m=R_{\perp}^2b^2_{\perp}(1-z^2/R_\parallel^2b^2)$. 
The averaged speed of sound reads 
\begin{equation}
\bar c_n^2(z,t)= \frac{g}{C_n b_\perp^2}
\int_{r<r_m} d^2r \rho_0 
\phi^2_n , 
\end{equation}
and the (space and time dependent) effective mass term is, for
a given radial mode, obtained to be
\begin{equation}
M_n^2(z,t)=\frac{g}{C_n b_\perp^2} \int_{r<r_m} d^2r \rho_0 
[\partial_r\phi_n 
]^2. \label{Mass} 
\end{equation}
The phonon branch of the excitations corresponds to the gapless $n=0$ part of the spectrum in 
Eq.\,(\ref{zarembadisprel}). In this case the  radial wavefunction $\phi_0$ does not depend on 
the radial variable $r$ \cite{Zaremba}, and the mass term vanishes, 
$M_0=0$.  
We then obtain the following expressions, 
\begin{equation}
C_0(z)=\pi R_{\perp}^2 \left(1-\frac{z^2}{b^2 R_\parallel^2}\right),
\label{C0z}
\end{equation}
and for the $z$ dependent speed of sound ($\bar c \equiv \bar c_{n=0}$):   
\begin{equation}
\bar c^2(z,t)= \frac{c^2_0}{b_{\perp}^2b}
\left(1-\frac{z^2}{b^2R_\parallel^2}\right).
\label{cbarz}
\end{equation}
We will see below that we need these expressions in the limit 
$z\rightarrow 0$ only, because only in this limit we get 
the desired exact mapping of the phonon field to a quantum field
propagating in a 1+1D curved space-time.

\section{The 1+1D de Sitter metric in the condensate center}
We identify the action (\ref{1daction}) with the action of a minimally
coupled scalar field in 1+1D, according to Eq.\,(\ref{action}).
Such an identification is possible only close to the center
$z=0$, as will now show. 
The {contravariant} 1+1D metric may generally be written as 
\cite{Matt}
\begin{equation}
{\sf g}^{\mu\nu}= \frac{1}{A_c c^2}  \left( \begin{array}{cc} -1 & -v_z \\ -v_z & c^2 -v_z^2 
\end{array} \right),
\end{equation}
where $A_c=A_c(x^\mu)$ is a conformal factor and $c=\bar c (z=0)$.
Inverting this expression, we get a covariant metric of the Painleve-Gullstrand form \cite{PG}, 
\begin{equation}
{\sf g}_{\mu\nu}= A_c \left( \begin{array}{cc} -(c^2-v_z^2) & -v_z \\ -v_z & 1
\end{array} \right). \label{gdown} 
\end{equation}
The term $\sqrt{-{\sf g}} {\sf g}^{\mu\nu}$ contained in the action
(\ref{action}) gives the familiar conformal invariance in a 1+1D space-time, 
i.e., the conformal factor $A_c$ drops out from the action and thus does 
not influence the (classical) equations of motion.
We therefore leave out $A_c$ in the formulae to follow, but it 
needs to be borne in mind that the metric elements 
are defined always up to the factor $A_c$. 

The actions (\ref{action}) and (\ref{1daction}) can be made consistent if 
we renormalize the phase field according to $\Phi = Z \tilde \Phi $ 
{\em and}  require that 
\begin{equation}
\frac{b_\perp^2 C_0 (0)}g Z^2 
= \frac1{\bar c} \label{ZC0}
\end{equation} 
holds.  
The factor $Z$ does not influence the equation of motion, 
but does crucially influence the strength of detector response (see section 
\ref{deSitterdetect} below).
In other words, it renormalizes the coupling of our ``relativistic field'' 
$\Phi$ to the detector. 
More explicitly, Eq.\,(\ref{ZC0}) leads to  
\begin{equation}
\frac{b_\perp}{{\sqrt b}}= 8 \sqrt{\frac{\pi}2} 
\frac1{Z^2} \sqrt{\rho_m a_s^3} 
\left(\frac{\omega_\perp}\mu\right)^2 \equiv B 
= {\rm const}. 
\label{cond1}
\end{equation} 
According to the above relation, 
we have to impose that the expansion of the cigar 
in the perpendicular direction proceeds like the square root
of the expansion in the axial direction. The constant quantity $B$ can be 
fixed externally (by the experimentalist), choosing the expansion of 
the cloud appropriately by adjusting the time dependence of the 
trapping frequencies $\omega_\parallel(t)$ and 
$\omega_\perp(t)$, as prescribed by the scaling equations
\cite{Scaling,0303063}
\begin{eqnarray}
\ddot b + \omega^2_\parallel (t) b =  
\frac{\omega^2_\parallel}{b_\perp^2 b^2} = 
\frac{\omega^2_\parallel}{B^2b^{3}} ,\qquad  
\ddot b_\perp + \omega^2_\perp (t) b_\perp   =   
\frac{\omega^2_\perp}{b_\perp^3 b} = 
\frac{\omega^2_\perp}{B^3b^{5/2}}.
\label{ScalingEqs}
\end{eqnarray}

Since both $C_0$ and $\bar c$ depend on $z$, Eqs.\,\eqref{C0z} and \eqref{cbarz},   
an effective space-time metric for the axial phonons can be obtained 
only close to the center of the cigar-shaped condensate cloud. 
This is related to our averaging over the physical
perpendicular direction, and does not arise if the excitations are 
considered in a higher-dimensional situation, where this identification
is possible globally. The reason for the 1+1D geometry is of practical origin: It implies that during
expansion, the coupling $g$ can be chosen independent of time to reproduce the de Sitter metric, which 
is in stark contrast to the 2+1D and 3+1D case, where the interaction needs to increase exponentially
for a de Sitter effective space-time to be obtained \cite{0303063}.
Also note that the action (\ref{1daction}) 
does not contain a curvature scalar contribution 
of the form $\propto \chi_n^2 \sqrt{-{\sf g}}\, {\sf R} 
[{\sf g_{\mu\nu}}(x)]$, i.e., that it only possesses trivial
conformal invariance \cite{Dotsenko}.   

We now impose, 
in addition, the requirement that the metric is identical 
to that of a 1+1D universe, 
with a metric of the form of the de Sitter metric in 3+1D 
\cite{Gibbons,de Sitter}.  
We first apply the transformation $c_0 d\tilde t = c(t) dt$ 
to the line element defined by (\ref{gdown}),  
connecting the laboratory time $t$ to the time variable $\tilde t$.
Defining $v_z/c = \sqrt \Lambda z = (B \dot b/c_0) z$ 
(note that the ``dot'' on $b$ and other quantities always refers
to laboratory time $t$),
this results, up to the conformal factor $A_c$, in the line element  
\begin{eqnarray}
d s^2 =-c_0^2(1- \Lambda z^2)d \tilde t^2 -2c_0 z \sqrt \Lambda d\tilde t 
dz+dz^2.
\end{eqnarray}
We then apply a second transformation 
$c_0 d\tau= c_0 {d\tilde t} +z\sqrt \Lambda dz/(1-\Lambda z^2)$, with a
constant $\Lambda$. 
We are thus led to the 1+1D de Sitter metric in the form 
\cite{Gibbons}
\begin{equation}
d s^2  = -c_0^2 \left(1-\Lambda  z^2 \right)  
d {\tau}^2 + \left({1-\Lambda z^2 }\right)^{-1} 
d z^2\,. 
\label{deSitterlineelement}
\end{equation}
The transformation between $t$ and the de Sitter time $\tau$ 
(on a constant $z$ detector, such that $d\tilde t = dt $), is given by
\begin{equation}
\frac{t}{t_0} = \exp[B \dot b \tau], \label{trafo} 
\end{equation} 
where the unit of lab time $t_0 \sim \omega_\parallel^{-1}$ 
is set by the initial conditions 
for the scaling variables $b$ and $b_\perp$.   
The temperature associated with the effective 
metric (\ref{deSitterlineelement}) is the Gibbons-Hawking 
temperature \cite{Gibbons} 
\begin{eqnarray}
T_{\rm dS} 
& = &  \frac{c_0}{2\pi } \sqrt{\Lambda} = \frac B{2\pi}  {\dot b}  .
\label{TdS}
\end{eqnarray}
The ``surface gravity'' on the horizon has the value 
$a_{\rm H} = c_0^2 \sqrt \Lambda = c_0 B \dot b$, and 
the stationary horizon(s) are located at the constant values
of the $z$ coordinate 
\begin{equation}
z=\pm {z_{\rm H}}= \pm {R_\parallel} \sqrt{\frac{\omega^2_\parallel}
{2\mu\Lambda}} . \label{rH}
\end{equation}
Combining (\ref{TdS}) and (\ref{rH}), we see that $z_{\rm H}/R_\parallel$
is small if $\omega_\parallel /T_{\rm dS}\ll 4\pi $. Therefore, the  
de Sitter temperature needs to be at least of the order of
$\omega_\parallel$ for the horizon location(s) to be well inside the cloud. 
The latter condition then justifies neglecting the $z$ dependence 
in $C_0$ and $\bar c$ in Eq.\,(\ref{ZC0}).
Though there is no metric ``behind'' the horizon, i.e. at large $z$, 
this should not affect the low-energy behavior of the quantum vacuum
``outside'' the horizon, i.e. in the center of the cloud.

%

\section{Determining the quasiparticle content of the quantum field}
As stated, the particle content of a quantum field state depends on the observer 
\cite{unruh76,BirrellDavies,fulling}. To verify this statement, we need to make the notion of 
observer operationally precise in that we find a quasiparticle detector which couples 
to the quantum field in question and hence measures only the quasiparticles corresponding to excitations above  
the appropriate vacuum.
Specifically, to detect the Gibbons-Hawking effect in de Sitter 
space, one 
has to set up a detector which measures frequencies in units of the inverse
de Sitter time $\tau$, rather than in units of the inverse laboratory time $t$.
The de Sitter time interval 
$d\tau=dt/bB=dt/{\sqrt b}b_\perp$ 
can be measured by an ``Atomic Quantum Dot'' (AQD) 
\cite{PRL,PRD,AQD}.
The measured quanta can then, and only then, 
be accurately interpreted to be quasiparticles coming from a Gibbons-Hawking 
type process with a {constant} de Sitter temperature (\ref{TdS}). 
In addition, we use the tunability to other
time intervals feasible with our detector scheme, 
and contrast the thermal de Sitter result with what the detector ``sees'' if tuned 
to laboratory and ``adiabatic'' time intervals. We will see that such a tuning is possible in the quantum-optical 
context by temporal variations of the laser amplitudes coupling the hyperfine levels of the atom.

The AQD can be manufactured in an ultracold gas of atoms possessing two hyperfine
ground states $\alpha$ and $\beta$. The atoms in the state $\alpha$ 
represent the superfluid cigar, and are used to model the expanding universe background on which the
phononic quasiparticles propagate. 
The AQD itself is formed by trapping atoms in the state $\beta$ in a 
tightly confining optical potential
$V_{\rm opt}$. The interaction of atoms in the two internal levels is
described by a set of coupling parameters $g_{cd} = 
4\pi a_{cd}$ ($c,d =\{\alpha,\beta\}$),
where $a_{cd}$ are the $s$-wave scattering
lengths characterizing short-range intra- and inter-species collisions;
$g_{\alpha\alpha}\equiv g$, $a_{\alpha\alpha} \equiv a_s$.
The on-site repulsion between the atoms $\beta$ 
in the dot is $U\sim g_{\beta \beta}/l^{3}$,
where $l$ is the characteristic size of the ground state wavefunction
of atoms $\beta$ localized in $V_{\rm opt}$. In the following, we consider the collisional
blockade limit of large $U>0$, where only one atom of type
$\beta$ can be trapped in the dot. This assumes that the double-occupancy gap $U$ is larger
than all other relevant frequency scales in the dynamics of both
the AQD and the expanding superfluid. 
As a result, the collective coordinate of the AQD is modeled by a 
pseudo-spin-$1/2$, with 
spin-up/spin-down state corresponding to occupation by a single/no atom
in hyperfine state $\beta$.

We first describe the AQD response to the condensate
fluctuations in the Lagrangian formalism, most familiar in a 
field theoretical context.
The detector Lagrangian takes the form 
\begin{multline}
L_{\rm AQD}  = 
i \left(\frac d{dt}{\eta^*} \right) 
\eta -\left[-\Delta+g_{\alpha\beta} (\rho_0(0,t) +\delta\rho)
\right]\eta^* \eta \\ 
- \Omega \sqrt{\rho_0(0,t) l^3}
\left( 
\exp\left[-i \int_0^t g \rho_0 (0,t') dt' + i\delta\phi\right]
\eta^*  
\label{coupling}
+ \exp\left[i \int_0^t g \rho_0 (0,t') dt' - i\delta\phi\right]\eta
\right).
\end{multline}
Here, $\Delta$ is the detuning of the laser light 
from resonance, $ 
\rho_0(z=0,t)$ is the central mean-field part of the bath density, 
and $l$ is the size of the 
AQD ground state wave function. 
The detector variable $\eta$ is an anticommuting Grassmann variable
representing the pseudo-spin degree of freedom of the AQD.  
The second and third lines represent the coupling of the AQD to the 
surrounding superfluid, where $\delta\phi$ and $\delta\rho$
are the fluctuating parts of the condensate phase and density 
at $z=0$, respectively.
The laser intensity and the effective transition matrix element combine
into the Rabi frequency $\Omega$; below we will make use of the fact that 
$\Omega$ can easily experimentally be changed as a function of 
laboratory time $t$, by changing the laser 
intensity with $t$. 

To simplify (\ref{coupling}), we use the canonical transformation 
\begin{eqnarray}
\eta & \rightarrow &  
\bar \eta \exp\left[-i \int_0^t g \rho_0 (0,t') dt' 
+ i\delta\phi\right]. \label{etatrafo} 
\end{eqnarray}
The above transformation amounts 
to absorbing the superfluid's chemical potential and 
the fluctuating phase  $\delta\phi$ into the wave function of the AQD, and
does not change the occupation numbers of the two AQD states.  
The transformation (\ref{etatrafo}) gives the detector Lagrangian the form 
\begin{eqnarray}
L_{\rm AQD} & = &  i 
\left(\frac{d}{dt} {{\bar \eta}^*} \right) 
\bar \eta 
- \Omega \sqrt{\rho_0(0,t) l^3}
\left(\bar \eta + \bar \eta^*\right) \label{Lbareta}  
-\left[-\Delta+(g_{\alpha\beta}-g)\rho_0 (0,t)
+g_{\alpha\beta} \delta \rho 
+ \frac d{dt} \delta\phi
\right] \bar \eta^* \bar \eta\,. \nonumber
\end{eqnarray}
The laser coupling (second term in the first line)
scales as $b^{-1/2} b^{-1}_\perp$, and hence like the de Sitter time 
interval in units of the laboratory time interval, $d\tau/dt$.
We suggest to operate the detector at the time dependent detuning 
$\Delta (t) =(g_{\alpha\beta}-g)\rho_{0}(0,t)=
(g_{\alpha\beta}-g)\rho_m/(\dot b^2 B^2 t^2)$, 
which then leads to a vanishing of the first 
two terms in the square brackets of (\ref{Lbareta}).
 
We now introduce back the wave function of the AQD stemming from 
a Hamiltonian formulation, $\psi
= \psi_\beta |\beta\rangle +\psi_\alpha |\alpha\rangle $. 
An ``effective Rabi frequency'' may be defined to be 
$\omega_0=2\Omega \sqrt{\rho_m l^ 3}$;  at the detuning compensated
point, we then obtain a  
simple set of coupled equations for the AQD amplitudes
\begin{equation}
i\frac{d \psi_\beta}{d\tau}=\frac{\omega_0}2 \psi_\alpha + \delta V \psi_\beta,\,\,\,\,\qquad 
i\frac{d \psi_\alpha}{d\tau}=\frac{\omega_0}2 \psi_\beta,\label{deSitterEvol}
\end{equation}
where $\tau$ {\em is the de Sitter time.} 

We have thus shown that the detector Eqs.\,(\ref{deSitterEvol}) are  
natural evolution equations in de Sitter time $\tau$, if the Rabi 
frequency $\Omega$ is chosen to be a constant, independent of laboratory time $t$.
We will see in sections \ref{detectlab} and \ref{detectMinkowski} 
 that, adjusting $\Omega$ in a certain
time dependent manner, within the same detector scheme, we can 
reproduce time intervals associated to various other effective space-times.

The coupling of the AQD to fluctuations 
in the superfluid is described by the perturbation potential
\begin{equation} 
\delta V (\tau) = \left(g_{\alpha\beta}-g \right)B b(\tau) \delta \rho(\tau).
\end{equation}
Neglecting the fluctuations in the superfluid,
the level separation implied by (\ref{deSitterEvol})
is $\omega_{0}$, and the eigenfunctions of the dressed two level
system are $|\pm\rangle= (|\alpha\rangle \pm |\beta\rangle)/\sqrt 2$. 
The quantity $\omega_0$ therefore 
plays the role of a frequency standard of the
detector. By adjusting the value of the laser intensity, one can change 
$\omega_0$, and therefore probe the response of the detector for 
various phonon frequencies. 
Note that if $g_{\alpha\beta}$ is very close to  $g$,  
to obtain the correct perturbation 
potential, higher order terms in the density 
fluctuations have to be taken into account 
in the Rabi term of (\ref{Lbareta}).

To describe the detector response, we first have to solve the 
equations of motion \,(\ref{1daction}) for the phase fluctuations, and
then evaluate the conjugate density fluctuations. 
The equation of motion  $\delta S_0 /\delta \chi_0=0$  is, 
for time independent $B$, given by
\begin{equation}
B^2 b^2 \partial_t \left( b^2 \partial_t \chi_0 \right) - \frac1{C_0 (z_b)} 
\partial_{z_b} \left({\bar c^2(z_b) C_0 (z_b)}\partial_{z_b} 
\chi_0\right)  \label{chiEqmotion}
= 0,
\end{equation}
where $z_b = z/b$ is the scaling coordinate. Apart from the  
factor $C_0(z_b)$, stemming from averaging over the perpendicular direction,
this equation corresponds to the hydrodynamic equation of phase
fluctuations in inhomogeneous superfluids \cite{Stringari}.  
At $t\rightarrow -\infty$, the condensate is in equilibrium and the
quantum vacuum phase fluctuations close to the center of the condensate can 
be written for a given wavevector $k$ in the following form 
\begin{equation}
\hat \chi_0 =
\sqrt{\frac{g}{4C_0(0) R_\parallel \epsilon_{0,k}}}
\hat a_{k} \exp\left[
-i\epsilon_{0,k} t + ikz \right]
+ {\rm H.c.}  \qquad (t\rightarrow -\infty),
\end{equation}
where $\hat a_k,\hat a^\dagger_k$ are the annihilation and creation 
operators of a phonon. The intial quantum state of phonons is the 
ground state of the superfluid and is annihilated by the operators 
$\hat a_k$. 
With these initial conditions, the solution of (\ref{chiEqmotion}) is  
\begin{eqnarray}
\hat \chi_0 & = &
\sqrt{\frac{g}{4C_0(0) R_\parallel \epsilon_{0,k}}}
\hat a_{k} \exp\!\left[
-i\int_0^t \frac{dt' \epsilon_{0,k}}{Bb^2}+ ikz_b 
\right]
+ {\rm H.c.} 
\label{chi0}
\end{eqnarray}
The solution for the density fluctuations, which are
 in a superfluid canonically 
conjugate to the phase fluctuations, $\delta\hat\rho= -\partial_t \hat\chi_0/g$ and 
$[\delta\hat\rho({\bm r}),\hat\chi_0({\bm r}')]=i\delta({\bm r}-{\bm r}')/(Bb^2)$ \cite{commutator}, 
therefore is 
\begin{eqnarray}
\delta\hat\rho & = & 
 -\sqrt{\frac{1}{4g C_0(0) R_\parallel \epsilon_{0,k}}} 
\frac{\partial}{\partial t}\! \left( 
\hat a_{k}\exp\!\left[
-i\int_0^t \frac{dt' \epsilon_{0,k}}{Bb^2}+ ikz_b \right] \right) 
+ {\rm H.c.} \label{deltarho} 
\end{eqnarray}
The Eqs.\,\,(\ref{chi0}) and (\ref{deltarho}) completely characterize
the evolution of the quantized hydrodynamic 
condensate fluctuations when the transverse modes are not excited. 
Observe that the evolution proceeds without frequency mixing in the adiabatic time 
interval defined by $d\tau_{\rm a}= dt/Bb^2$ 
(the ``scaling time'' interval $dt/B^2b^2$ defined in \cite{0303063} 
is proportional to this adiabatic time interval).  
Therefore, in the ``adiabatic basis,'' no frequency mixing occurs and thus no 
quasiparticle excitations are created, see also section \ref{detectMinkowski} below. 
This hints at a hidden (low energy)
symmetry,
in analogy to the (exact) 2+1D Lorentz group SO(2,1) for an isotropically
expanding BEC disk, discussed in \cite{PitaRosch}. 

\subsection{Quasiparticle detection in de Sitter time}  \label{deSitterdetect}
The coupling operator $\delta \hat V$ causes transitions between the 
dressed detector states $|+\rangle$ and $| -\rangle$ 
and thus can be used to effectively measure the quantum state of the phonons. 
We consider the detector response to fluctuations 
of $\hat \Psi$, by going beyond mean-field and 
using a perturbation theory in $\delta\hat V$.
There are two physically different situations. The detector 
is either at $t=0$ in its ground state,  
$(|\alpha\rangle + |\beta\rangle)/\sqrt 2$, 
or in its excited state, $(|\alpha\rangle - |\beta\rangle)/\sqrt 2$. 
We define $P_+$ and $P_-$ to be the probabilities that 
at late times $t$ the detector is excited respectively de-excited.
Using second order perturbation 
theory in $\delta\hat V$, we find that 
the transition probabilities for the detector may be written 
\begin{eqnarray}
P_{\pm} & = & \sum_k \frac{g\epsilon_{0,k}}{4R_\parallel C_0(0)} 
\left(\frac{g_{\alpha\beta}}{g}-1\right)^2 {B}^2 \left|T_\pm\right|^2 ,
\label{Ppm}
\end{eqnarray} 
where the absolute square of the transition matrix element is given by 
\begin{equation}
|T_\pm|^2 =  \left|\int_{0}^\infty \frac{d\tau}{b(\tau)} 
\exp\left[\pm i\epsilon_{0,k}\int_0^\tau 
\frac{d\tau'}{b(\tau')} + i \omega_0\tau \right]\right|^2 . \label{T(tau)}
\end{equation} 
Calculating the integrals, we obtain 
\begin{eqnarray}
P_{\pm} & = & J \left(\frac {g_{\alpha\beta}}g -1 \right)^2 {B}^2  
\frac{g\pi}{2B \dot b R_\parallel C_0(0)} 
\left\{\begin{array}{c} n_{\rm B}
\\
1+n_{\rm B} 
\end{array} ,\right.
\end{eqnarray}
where a (formally divergent) sum is contained in the factor 
\begin{equation}
J= \sum_k \frac{\omega_0}{\epsilon_{0,k}},\label{J}
\end{equation}
and we obtain Bose-Einstein distribution functions at boson energy 
$\omega_0$, containing the de Sitter temperature (\ref{TdS}):
\begin{equation}
n_{\rm B} = \frac{1}{\exp[\omega_0/T_{\rm dS}]-1}.
\end{equation} 
We conclude that an expansion of the condensate in $z$ direction, with 
a constant rate faster than the harmonic 
trap oscillation frequency in that direction, 
gives an effective de Sitter space-time, characterized by a thermal distribution at the de Sitter temperature
$T_{\rm dS}$, as measured by the AQD in its natural time interval $d\tau$.

We now show that $J$ in \eqref{J} is proportional to the total {\em de Sitter} time 
of observation, so that the probability per unit time
is a finite quantity \cite{Menshikov}. 
At late times, the detector measures phonon quanta coming, relative
to its space-time perspective, 
from close to the horizon, at a distance
$\delta z = z_{\rm H}-z \ll z_{\rm H} =\Lambda^{-1/2}$. 
The trajectory of such a phonon in the coordinates of the de Sitter metric 
(\ref{deSitterlineelement}), at late times $\tau$, is given by 
\begin{equation}
\ln \left[\frac{z_{\rm H}}{\delta z}\right] = 2\sqrt\Lambda c_0 \tau.
\end{equation}
This implies that the central AQD detector 
measures quanta which originated at the horizon with the large 
blueshifted frequency
\begin{equation}
\epsilon_{0,k} = \frac{\omega_0}{\sqrt 2 \Lambda^{1/4}\delta z^{1/2}} =
\frac{\omega_0}{\sqrt 2}\exp[c_0\tau \sqrt\Lambda].
\end{equation}
Making use of the above equation, we rewrite the summation over 
$k$ in (\ref{J}) as an integral over detector time:
\begin{eqnarray}
J=\sum_k \frac{\omega_0}{\epsilon_{0,k}} &  = &
\frac{R_\parallel\omega_0}{\pi c_0}\int \frac{d\epsilon_{0,k}}{\epsilon_{0,k}}
= \frac{R_\parallel\omega_0}{\pi c_0} \sqrt\Lambda c_0 \int d\tau \nonumber\\
& = &  \frac{R_\parallel\omega_0}{\pi c_0} B\dot b \tau \,.
\end{eqnarray} 
Therefore, 
the probabilities per unit detector time (de Sitter time) read, where 
upper/lower entries refer to $P_+$/$P_-$, respectively:
\begin{eqnarray}
\frac{d P_{\pm}}{d\tau} 
& = & \left(\frac{g_{\alpha\beta}}g -1 \right)^2 {B}^2   
\frac{g \omega_0}{2C_0(0) c_0} \left\{
 \begin{array}{c} n_{\rm B} 
\\
1+n_{\rm B}
\end{array} 
\right. .\label{PdtaudS}
\end{eqnarray}
They are finite quantities in the limit that $\tau\rightarrow \infty$. 
In laboratory time, the transition probabilities evolve according to 
\begin{eqnarray}
P_{\pm} (t) 
& = & P_0 
\frac{\omega_0}{T_{\rm dS}} \ln \left[\frac t{t_0} \right] 
\left\{ \begin{array}{c} n_{\rm B} 
\\
1+n_{\rm B}
\end{array} ,
\right. \label{Ppm(t)}
\end{eqnarray}
where, from relation (\ref{ZC0}), $P_0 = Z^2 [({g_{\alpha\beta}}/g-1)B]^2/2 $.
We see 
that the detector response is, as it should be, proportional to $Z^2$, 
the square of the renormalization factor of the phase fluctuation field.

The absorption and emission coefficients ${d P_{\pm}}/{d\tau} $
satisfy Einsteinian relations. Therefore, the detector approaches
thermal equilibrium at a temperature $T_{\rm dS}$ on a time scale 
proportional to $Z^{-2} \omega_0^{-1}$. 
Remarkably, our de Sitter 
AQD detector thus measures a stationary thermal 
spectrum, even though its  condensed matter 
background, with laboratory time $t$, is in a 
highly nonstationary motional state. 
Since $Z^2 \propto 
\sqrt{\rho_m a_s^3} \left({\omega_\perp}/\mu\right)^2$, 
not-too-dilute condensates with $\omega_\perp \sim \mu $ 
(i.e., close to the quasi-1D r\'egime \cite{Gorlitz})
are most suitable to observe the Gibbons-Hawking effect. 

The verification of the fact that a thermal detector 
state has been established
proceeds by the fact that the two hyperfine states $\alpha$ and $\beta$ are
spectroscopically different states of the same atom, easily  
detectable by modern quantum optical technology.
When the optical potential is switched on, the atoms are
in the empty $\alpha$ state originally, 
which is an equal-weight superposition of 
$|+\rangle = (|\alpha\rangle + |\beta\rangle)/\sqrt 2$
and $|-\rangle = (|\alpha\rangle - |\beta\rangle)/\sqrt 2$.
The thermalization due to the Gibbons-Hawking effect takes place 
in the dressed state basis consisting of the two {\em detector states}, 
i.e. of the states $|+\rangle$ and $|-\rangle$, 
on a time scale given by the quantities $P_\pm$ in Eq.\,(\ref{Ppm(t)}). 
For the laboratory observer, the Gibbons-Hawking thermal 
state will thus appear to cause damping of the Rabi oscillations 
on the thermalization timescale, i.e., friction on 
the coherent oscillating motion between the two 
detector states occurs, due to the 
thermal phonon bath perceived by the detector. 
 The occupation of the {detector states} 
can be measured directly using atomic interferometry: A $\pi/2$-pulse brings
one of them into the filled 
($\beta$) and the other into the empty ($\alpha$) state.
To increase the signal to noise ratio, one could conceive of 
manufacturing a small 
array of AQDs in a sufficiently large cigar-shaped host superfluid, 
and monitor the total population of $\beta$ atoms in this array.

\subsection{Detection of quasiparticles in laboratory time} \label{detectlab}
We contrast the above calculation with the response 
the AQD detector would see if tuned to laboratory time. 
This can be realized if we let $\Omega \propto t$, such that the quantity 
$\Omega \sqrt{\rho_0(0,t)}= \Omega \sqrt{\rho_m} / (B\dot b t) 
= {\rm const.}$, in the Rabi term on the right-hand side 
of (\ref{Lbareta}), is time 
independent. The detector has, therefore, the laboratory time interval $dt$ in the equations 
for the occupation amplitude with this choice for $\Omega = \Omega(t)$. The Painlev\'e-Gullstrand metric
(\ref{gdown}) in pure laboratory frame variables, assuming
$B^2\dot b^2 \gg  \omega_\parallel^2 /b^2$
like in the derivation of the de Sitter metric 
(\ref{deSitterlineelement}), reads 
\begin{equation}
ds^2 = -\frac{c_0^2}{B^2 \dot b^2 t^2}\left(1- \Lambda z^2\right)dt^2
-\frac{2z}t dz dt +dz^2. \label{labframemetric}
\end{equation}
The metric  (\ref{labframemetric}) 
is asymptotically, for large $t$, becoming that of 
Galilei-invariant ordinary Newtonian laboratory space, i.e. is just measuring
length along the $z$ direction, because the speed of sound
in the ever more dilute gas decreases like $1/t$ and the 
``phonon ether'' becomes increasingly less stiff.
 
The transition probabilities for absorption respectively
emission are now given by 
\begin{eqnarray}
{\tilde P}_{\pm} =  \frac{g\epsilon_{0,k}}{4R_\parallel C_0(0)} 
\left({g_{\alpha\beta}}/{g}-1\right)^2 |\tilde T_\pm |^2,
\end{eqnarray}
where the matrix elements are, cf. Eq.\,(\ref{T(tau)}), 
\begin{equation}
|\tilde T_\pm|^2 =  \left|\int_{-\infty}^\infty \frac{dt}{Bb^2} 
\exp\left[\pm i\epsilon_{0,k}\int_{-\infty}^\infty 
\frac{dt'}{Bb^2} + i \omega_0 t \right]\right|^2 .
\end{equation} 
Substituting the adiabatic time interval 
$d\tau_{\rm a}  = dt/ (B\dot b^2 t^2)$ leads for large $t$ to 
\begin{equation}
\tau_{\rm a} = \tau_{0s} -(B\dot b^2 t)^{-1}, 
\end{equation}
where $\tau_{0{\rm a}}= \int_{-\infty}^{+\infty}
dt (B\dot b^2 t^2)^{-1}$. The transformation to adiabatic time
maps $t\in [-\infty, +\infty]$ onto $\tau_{\rm a} \in 
[-\infty,\tau_{0{\rm a}}]$
and, by further substituting  
$y = \epsilon_{0,k} (\tau_{\rm a}-\tau_{0{\rm a}})$, we have 
\begin{equation}
|\tilde T_\pm|^2 = \frac{1}{\epsilon_{0k}^2} 
\left|\int_{0}^\infty dy  
\exp\left[i\left( y
\mp \frac{\omega_0 \epsilon_{0k}}{B\dot b^2} \frac 1y
\right) \right]\right|^2 .
\end{equation} 
The integral is 
a linear combination of Bessel functions.
To test its convergence properties, 
we are specifically interested in the large $\epsilon_{0k}$ 
limit. Performing a stationary phase approximation for large
$A= \pm {\omega_0 \epsilon_{0k}}/{B\dot b^2}$, we have for 
positive $A$ (absorption) that the integral above becomes
$J(A)=(\pi \sqrt A)^{1/2} \exp[-2\sqrt A]$ and 
for negative $A$ (emission) $J(A) = (\pi \sqrt{|A|})^{1/2}$. 
The final result for the laboratory time transition  probabilities then is 
\begin{eqnarray}
{\tilde P}_{\pm} 
&= &\left(\frac{g_{\alpha\beta}}{g}-1\right)^2  \sqrt{2\pi}  
\sqrt{\rho_m a_s^3} \left(\frac{\omega_\perp}\mu \right)^2 
\int_0^{E_{\rm Pl}}\! d\epsilon_{0k} 
\sqrt{\frac{\omega_0}{\epsilon_{0k} B \dot b^2 }} 
\times \left\{
\begin{array}{c} \exp\left[-4\sqrt{\frac{\omega_0\epsilon_{0k}}{B\dot b^2}} \,
\right]
\\ 1 
\end{array}  
\right.  \nonumber\\
&= &\left(\frac{g_{\alpha\beta}}{g}-1\right)^2  
\!\sqrt{{2\pi \rho_m a_s^3}} 
\left(\frac{\omega_\perp}\mu \right)^2 
\times \left\{
\begin{array}{c}\! \simeq \frac12 \\ 
\\\!  \sqrt{\frac{4E_{\rm Pl} \omega_0}{B \dot b^2}}
\end{array}  
\right. ,\nonumber\\
\label{labtimeP} 
\end{eqnarray}     
where $E_{\rm Pl} \sim \mu$ is the ultraviolet cutoff in the integral for 
the emission probability ${\tilde P}_-$ , the ``Planck'' scale of the 
superfluid. 
Because of the convergence of the absorption integral for ${\tilde P}_+$, the
total number of quasiparticles detected remains finite, and there are thus no
quasiparticles detected by the laboratory-frame-tuned detector at late times. 
This is in contrast to the de Sitter detector, which 
according to (\ref{PdtaudS}) detects quasiparticles in a stationary thermal state.

\subsection{Adiabatic ``Minkowski'' basis with no quasiparticles detected} \label{detectMinkowski}

There is a detector setting which corresponds to a detector at rest
in the Minkowski vacuum.
This setting is represented by the adiabatic basis, 
with time interval defined by $d\tau_{\rm a} = dt /Bb^2$, 
realizable with the AQD by decreasing the Rabi frequency linearly in time, $\Omega \propto 1/t$. 
Then, no quasiparticles whatsoever are detected at any instant, i.e., no
frequency mixing of the positive and 
negative frequency parts of (\ref{deltarho}) does take place. 
The associated space-time interval 
\begin{equation} 
ds^2 = b^2[-c_0^2 d\tau_{\rm a}^2 +dz_b^2]
\end{equation}
is simply that of (conformally) flat 
Minkowski space in the spatial scaling coordinate 
$z_b$ and adiabatic time coordinate $\tau_{\rm a}$. 
Specifically note here that the adiabatic time interval is {\em not} equal to laboratory time, $d\tau_{\rm a}\propto dt/t^2$. The detector in its effective Minkowski space hence measures, using this proper time interval,  that dissipation-free expansion of  the BEC cigar takes place. The ``absolute'' time of the experimentalist in the lab frame can thus be different from the proper time interval of the detector in the sonic Minkowski frame.
The ``internal'' observer therefore can ascertain independently from the lab experimentalist that indeed dissipation-free expansion of the background takes place and no real quasiparticles are created.

\section{concluding considerations}

We have demonstrated that the dependence of quasiparticle detection on the motional state of the detector 
and on the way the detector couples to the quantum field can be made experimentally manifest in a tabletop setup.  
An {atomic quantum dot} placed at the center of a linearly 
expanding cigar-shaped Bose-Einstein condensate has a proper time interval which can be tuned using quantum optical means like the intensity of  a laser.  
The proposed quantum optics experiment can thus confirm in the effective curved space-time setting of analogue gravity that indeed ``A particle detector will react to states which have positive frequency with respect to the detector's proper time, not with respect to any universal time \cite{unruh76}.''

The equilibration time scale of the detector, and thus the time scale
on which the Rabi oscillations between the detector states are damped
out, is set by the detector frequency standard (the level spacing) 
 $\omega_0$, and by the renormalization factor $Z$:
\begin{equation}
\tau_{\rm equil} = 
Z^{-2} \omega_0^{-1} 
\propto ({\rho_m a_s^3})^{-1/2} 
\left(\mu/{\omega_\perp}\right)^2 \omega_0^{-1} .
\label{tauequil}
\end{equation}
The renormalization factor $Z$ contained in 
(\ref{cond1}) determines the equilibration rapidity because
it physically expresses the strength of detector-field coupling, resulting in the transition probabilities
\eqref{Ppm(t)}. 
It is related to the initial diluteness parameter 
$D_p(0) = ({\rho_m a_s^3})^{1/2}$ of the
Bose-Einstein condensate and to the ratio $\mu/\omega_\perp$, which determines
inasmuch the system is effectively one-dimensional, 
$Z^2 \propto D_p(0) (\omega_\perp/\mu)^2$. 
To obtain sufficiently fast equilibration, the condensate thus 
has to be initially not too dilute as well as close to the quasi-1D
r\'egime, for which the transverse harmonic oscillator energy scale 
is of order the energy per particle, $\mu \sim \omega_\perp$. These two conditions 
have another important implication. 
The ratio of the instantaneous coherence length 
$\xi_c (t) = (8\pi \rho_0(0,t) a_s)^{-1/2} \propto t $ 
and the location of the horizons $z= z_{\rm H}= \pm \Lambda^{-1/2}$, 
which are {\em stationary} in the present setup, 
has to remain less than unity 
within the equilibration time scale.
If this is not the case, the coherence length, which 
plays the role of the analogue Planck scale (which is laboratory time
dependent here), exceeds the length scale of the horizon at equilibration, 
and the concept of ``relativistic'' phonons
propagating on a fixed curved space-time background with local Lorentz symmetry 
becomes invalid. 
The ratio $\xi_c (t) /z_H$ at the lab equilibration time scale 
$t= t_{\rm equil} = t_0 \exp[2\pi (T_{\rm dS}/\omega_0) 
\left(\mu/{\omega_\perp}\right)^2 D^{-1}_p(0)]$ 
following from the de Sitter equilibration time in Eq.\,(\ref{tauequil}), 
expressed in experimentally relevant scaled parameters, reads 
\begin{equation}
\frac{\xi_c (t_{\rm equil})}{z_{\rm H}} = 
\frac{\pi t_0 T^2_{\rm dS}}{\rho_m a_s} 
\exp\left[2\pi \frac{T_{\rm dS}}{\omega_0}  
\left(\frac{\mu}{\omega_\perp}\right)^2 \frac1{D_p(0)}\right].
\end{equation}
We see that this ratio changes exponentially with both the
initial diluteness parameter $D_p(0)$ 
and the quasi-1D parameter $\mu/\omega_\perp$.
In most currently realized Bose-Einstein condensates, with comparatively long lifetime, 
the diluteness parameter $D_p \sim 10^{-2}$. Here, we initially need 
$D_p (0)\lesssim O(1)$ to have the condition 
${\xi_c (t_{\rm equil})}/{z_{\rm H}}< 1$ fulfilled, 
assuming a reasonably large value of the de Sitter temperature 
$T_{\rm dS}$.  
Though the condensate has to be {\em initially} 
quite dense, it is to be stressed that the central density 
decays like $t^{-2}$ during expansion. Therefore, the 
rate of three-body recombination losses quickly decreases
during the expansion of the gas, and the initially relatively 
dense Bose-Einstein condensate, which would rapidly decay
if left with a $D_p$ close to unity, 
can live sufficiently long, the total rate of three-body 
losses decreasing like $\rho^2_0(0,t) \propto t^{-4}$. 

The gas parameter $D_p$ plays the role of a loop expansion parameter, which measures
the extent to which a mean-field theory applies \cite{LHY}. 
The Gibbons-Hawking effect in the BEC 
is therefore intrinsically quantum because the signal measured by the detector, Eq.\,\eqref{Ppm(t)}, 
as already stated above, contains this ``dimensionless Planck constant''  $D_p(0)$ at the initial laboratory time. 
When $D_p$ becomes large, of order unity, the 
mean-field theory employed in the above considerations strictly speaking breaks down.
Because the gas parameter enters the magnitude of the signal, a sufficiently large signal-to-noise ratio,
together with rapid equilibration of the detector, can however be achieved only by using initially dense clouds with 
strong interparticle interactions and relatively large gaseous parameter. The required initial state can be prepared by using a sufficiently rapid external magnetic field sweep to a Feshbach resonance to increase $a_s$ for 
an already existing BEC, 
and shortly thereafter (to avoid large three-body losses) expand, according to the required temporal behavior of scaling parameters in Eq.\,\eqref{ScalingEqs}. 

Considering the nonclassicality aspect of the Gibbons-Hawking effect from a different angle, 
the phononic quasiparticles of the superfluid can be regarded as
non-interacting only in a first approximation in $D_p \ll 1$.
Self-interaction between the phonons, induced by larger values of the gas parameter, can lead to 
decoherence and the relaxation of the phonon subsystem. 
The same line of reasoning applies to the evolution
of interacting quantum fields in the expanding universe. The interactions between
quasiparticle excitations in the BEC induced by them scattering off each other 
can therefore be related to decoherence processes 
in cosmological models of quantum field propagation and particle production.
Finally, as already mentioned in section \ref{ExpandBEC}, the Einsteinian part of the effective action 
becomes increasingly important for larger $D_p$, so that the combined effects of backaction of quantum fluctuations on the sonic metric and of decoherence due to phonon-phonon interactions may lead to increasingly 
rich behavior of the quantum field evolution.

\acknowledgments
The present contribution was supported by the 
NRF of Korea, grant Nos. 2010-0013103 and 2011-0029541. 




\begin{thebibliography}{9999}

\newfont{\cyr}{wncyr10}

\bibitem{unruh76} W.\,G. Unruh, 
Phys. Rev. D {\bf 14}, 870 (1976).
\bibitem{BirrellDavies}  P.\,C.\,W. Davies, J. Phys. A. {\bf 8}, 609 (1975);
N.\,D. Birrell and P.\,C.\,W. Davies, 
{\em Quantum Fields in Curved Space} (Cambridge University Press, Cambridge, 
England, 1984). 
\bibitem{fulling} S.\,A. Fulling,
Phys. Rev. D {\bf 7}, 2850 (1973).
\bibitem{Luis} L.\,C.\,B. Crispino, A. Higuchi, and G.\,E.\,A. Matsas,
Rev. Mod. Phys. {\bf 80}, 787 (2008). 
\bibitem{Yablonovitch} E. Yablonovitch,  Phys. Rev. Lett. {\bf 62}, 
1742 (1989).
\bibitem{chenUnruh} P. Chen and T. Tajima, 
Phys. Rev. Lett. {\bf 83}, 256 (1999). 
\bibitem{Habs} R. Sch\"utzhold, G. Schaller, and D. Habs, 
Phys. Rev. Lett. {\bf 97}, 121302 (2006). 
\bibitem{unruh} W.\,G. Unruh, Phys. Rev. Lett. {\bf 46}, 1351 (1981). An early precursor to Unruh's metric, derived from a different perspective, may be found in A. Trautman, ``Comparison of Newtonian
and Relativistic Theories of Space-Time'',  pp. 413--425 in {\em Perspectives in
Geometry and Relativity} 
(Indiana University Press, Bloomington, 1966).
\bibitem{BLVReview} The most comprehensive survey of Analogue Gravity,  containing some 700 citations,  may be found in its most current version in C. Barcel\'o, S. Liberati, and M. Visser, 
Living Rev. Relativity {\bf 14} (2011), 3.
\bibitem{Grisha} G.\,E. Volovik, 
{\it The Universe in a Helium Droplet} 
(Oxford University Press, Oxford, 2003).
\bibitem{GrishaTed} T.\,A. Jacobson and G.\,E.  Volovik, 
Phys. Rev. D {\bf 58}, 064021 (1998); JETP Lett. {\bf 68}, 874 (1998). 
\bibitem{Matt} M. Visser, Class. Quantum Grav. {\bf 15}, 1767 (1998).
\bibitem{SchuetzUnruh} R. Sch\"utzhold and W.\,G. Unruh, 
Phys. Rev. D {\bf 66}, 044019 (2002).
\bibitem{SlowLight} W.\,G. Unruh and R. Sch\"utzhold, 
Phys. Rev. D {\bf 68}, 024008 (2003).
\bibitem{PGRiemann} U.\,R. Fischer and M. Visser, 
Ann. Phys. (N.Y.) {\bf 304}, 22 (2003);  
Phys. Rev. Lett. {\bf 88,} 110201 (2002).
\bibitem{Wilczek} F. Wilczek, Physics Today, January 1999, pp. 11-13. 
\bibitem{CSM} C. Barcel\'o, S. Liberati, and M. Visser,
Class. Quantum Grav. {\bf 18}, 1137 (2001).
\bibitem{Picokelvin} A. E. Leanhardt {\it et al.}, 
Science {\bf 301}, 1513 (2003). 
\bibitem{BLVPRA} C. Barcel\'o, S. Liberati, and M. Visser,
Phys. Rev. A {\bf 68}, 053613 (2003).
\bibitem{0303063} P.\,O. Fedichev and U.\,R. Fischer,
Phys. Rev. A {\bf 69}, 033602 (2004).
\bibitem{Micha} M. Uhlmann, 
New J. Phys. {\bf 12}, 095016 (2010).
\bibitem{Prain} A. Prain, S. Fagnocchi, and S. Liberati,
Phys. Rev. D {\bf 82}, 105018 (2010).
\bibitem{Macher} J. Macher and R. Parentani, 
Phys. Rev. A {\bf 80}, 043601 (2009). 
\bibitem{Balbinot} R. Balbinot {\it et al.}, 
Phys. Rev. A {\bf 78}, 021603(R) (2008). 
\bibitem{Recati} 
A. Recati, N. Pavloff, and I. Carusotto,
Phys. Rev. A {\bf 80}, 043603 (2009).
\bibitem{MattHawking} M. Visser, Phys. Rev. Lett. {\bf 80}, 3436 (1998).
\bibitem{Garay} L.\,J. Garay, J.\,R. Anglin, J.\,I. Cirac, and 
P. Zoller, Phys. Rev. Lett. {\bf 85}, 4643 (2000).
\bibitem{Lahav} O. Lahav {\it et al.},
Phys. Rev. Lett. {\bf 105}, 240401 (2010). 
\bibitem{Gibbons} G.\,W. Gibbons and S.\,W. Hawking,  
Phys. Rev. D {\bf 15}, 2738 (1977).
\bibitem{de Sitter} W. de Sitter,  
Mon. Not. R. Astron. Soc. {\bf 78}, 3  (1917).
\bibitem{PRL} P.\,O. Fedichev and U.\,R. Fischer,   
Phys. Rev. Lett. {\bf 91}, 240407 (2003).
\bibitem{PRD} P.\,O. Fedichev and U.\,R. Fischer, 
Phys. Rev. D {\bf 69}, 064021 (2004).  
\bibitem{Retzker} A. Retzker, J.\,I. Cirac, M.\,B. Plenio, and B. Reznik,
Phys. Rev. Lett. {\bf 101}, 110402 (2008).
\bibitem{RalfPRL06}  R. Sch\"utzhold, 
Phys. Rev. Lett. {\bf 97}, 190405 (2006).
\bibitem{Menicucci} N.\,C. Menicucci, S.\,J. Olson, and G.\,J. Milburn,
New J. Phys. {\bf 12}, 095019 (2010).
\bibitem{Petersen} R. Sch\"utzhold {\it et al.}, 
Phys. Rev. Lett. {\bf 99}, 201301 (2007). 
\bibitem{Horstmann} B. Horstmann, B. Reznik, S. Fagnocchi, and  J.\,I. Cirac,
Phys. Rev. Lett. {\bf 104}, 250403 (2010).
\bibitem{BR} R. Sch\"utzhold, M. Uhlmann, Y. Xu, and U.\,R. Fischer,
Phys. Rev. D {\bf 72}, 105005  (2005); U.\,R. Fischer,  
Lect.\,\,Notes\, Phys. {\bf 718}, 93 (2007).
\bibitem{LiberatiPRL} S. Finazzi, S. Liberati, and L. Sindoni,  
Phys. Rev. Lett. {\bf 108}, 071101 (2012). 
\bibitem{MattMPL} The Einstein part of the action stems from integrating out quantum fluctuations at one-loop order
in the sense of Sakharov's induced gravity,
cf. M. Visser,
Mod. Phys. Lett. A {\bf 17},  977 (2002);  C. Barcel\'o, S. Liberati, and M. Visser,
Class. Quantum Grav. {\bf 18}, 3595 (2001).
\bibitem{LHY} 
T.\,D. Lee, K.\,W. Huang, and C.\,N. Yang, Phys. Rev. {\bf 106}, 1135 (1957).
\bibitem{Scaling} Yu. Kagan, E.\,L. Surkov,  and G.\,V. Shlyapnikov, 
Phys. Rev. A {\bf 54}, R1753 (1996); 
Y. Castin and R. Dum, Phys. Rev. Lett. {\bf 77}, 5315 (1996).
\bibitem{BaymPethick} G. Baym and C. J. Pethick, Phys. Rev. Lett. 
{\bf 76}, 6 (1996). 
\bibitem{PitaRosch} L. P. Pitaevski\v{\i} and A. Rosch,
Phys. Rev. A {\bf 55}, R853 (1997).
\bibitem{DalibardBreathing} 
F. Chevy {\it et al.}, 
Phys. Rev. Lett. {\bf 88}, 250402 (2002). 
\bibitem{Zaremba} E. Zaremba, Phys. Rev. A {\bf 57}, 518 (1998).
\bibitem{PG} P. Painlev\'e,
C. R. Hebd. 
Acad. Sci. (Paris) 
{\bf 173}, 677 (1921); 
A. Gullstrand, 
Arkiv. Mat. Astron. Fys. {\bf 16}, 1 (1922).
\bibitem{Dotsenko}  V.\,S. Dotsenko and V.\,A. Fateev,  
Nucl. Phys. B {\bf 240}, 312 (1984).
\bibitem{AQD} A. Recati {\it et al.}, 
Phys. Rev. Lett. {\bf 94}, 040404 (2005).
\bibitem{Stringari} S. Stringari, 
Phys. Rev. Lett. {\bf 77}, 2360 (1996). 
\bibitem{commutator} Observe that the laboratory frame commutator of density and phase fluctuations here receives an additional factor of $1/(Bb^2)$, as
compared with 
a stationary superfluid. 
\bibitem{Menshikov} L.\,I. Men'shikov and A.\,N. Pinzul, 
Phys. Usp. {\bf 38}, 1031 (1995).
\bibitem{Gorlitz} A. G\"orlitz {\it et al.},  
Phys. Rev. Lett. {\bf 87}, 130402 (2001).  
\end{thebibliography}
\end{document}